\documentclass[aps,prl,twocolumn,superscriptaddress,showpacs,floatfix]{revtex4-2}
\usepackage{color}
\usepackage{mathtools}
\usepackage[c]{esvect}

\usepackage{newunicodechar}

\usepackage{bm}        
\usepackage{amssymb}   
\usepackage{amssymb, amsthm, float, graphicx, amsfonts, dsfont}
\usepackage{amsmath} 

\usepackage[colorlinks=true,citecolor=blue,linkcolor=magenta, breaklinks=true]{hyperref}
\usepackage{graphicx}
\usepackage{braket}
\usepackage{natbib}
\usepackage [utf8]{inputenc}
\usepackage{siunitx}
\usepackage{blkarray, bigstrut}
\usepackage[dvipsnames]{xcolor}
\usepackage{comment}
\usepackage[utf8]{inputenc}
\usepackage{physics}
\usepackage{tikz}

\newcolumntype{P}[1]{>{\centering\arraybackslash}p{#1}}
\begin{document}

\title{Accessing quasi-flat \textit{f}-bands to harvest large Berry curvature in NdGaSi}

\author{Anyesh Saraswati}
\affiliation{S. N. Bose National Centre for Basic Sciences, Salt Lake City, Kolkata-700106, India}

\author{Jyotirmoy Sau}
\affiliation{S. N. Bose National Centre for Basic Sciences, Salt Lake City, Kolkata-700106, India}

\author{Vera Misheneva}
\affiliation{Helmholtz-Zentrum Berlin für Materialien und Energie, Albert-Einstein-Straße 15, 12489 Berlin, Germany}
\affiliation{Joint Laboratory “Functional Quantum Materials” at BESSY II, 12489 Berlin, Germany}

\author{Rui Lou} 
\affiliation{Helmholtz-Zentrum Berlin für Materialien und Energie, Albert-Einstein-Straße 15, 12489 Berlin, Germany}
\affiliation{Joint Laboratory “Functional Quantum Materials” at BESSY II, 12489 Berlin, Germany}
\affiliation{Leibniz Institute for Solid State and Materials Research, IFW Dresden, 01069 Dresden, Germany}

\author{Sudipta Chatterjee} 
\affiliation{S. N. Bose National Centre for Basic Sciences, Salt Lake City, Kolkata-700106, India}

\author{Sandip Kumar Kuila} 
\affiliation{Department of Chemistry, Indian Institute of Technology, Kharagpur-721302, India}

\author{Bibhas Ghanta} 
\affiliation{Solid State Physics Division, Bhabha Atomic Research Centre, Mumbai-400085, India}
\affiliation{Homi Bhabha National Institute, Anushaktinagar, Mumbai-400094, India}

\author{Anup Kumar Bera} 
\affiliation{Solid State Physics Division, Bhabha Atomic Research Centre, Mumbai-400085, India}
\affiliation{Homi Bhabha National Institute, Anushaktinagar, Mumbai-400094, India}

\author{Partha Pratim Jana} 
\affiliation{Department of Chemistry, Indian Institute of Technology, Kharagpur-721302, India}

\author{Alexander Fedorov}
\affiliation{Helmholtz-Zentrum Berlin für Materialien und Energie, Albert-Einstein-Straße 15, 12489 Berlin, Germany}
\affiliation{Joint Laboratory “Functional Quantum Materials” at BESSY II, 12489 Berlin, Germany}
\affiliation{Leibniz Institute for Solid State and Materials Research, IFW Dresden, 01069 Dresden, Germany}

\author{Setti Thirupathaiah}
\affiliation{S. N. Bose National Centre for Basic Sciences, Salt Lake City, Kolkata-700106, India}

\author{Manoranjan Kumar}\email{manoranjan.kumar@bose.res.in}
\affiliation{S. N. Bose National Centre for Basic Sciences, Salt Lake City, Kolkata-700106, India}

\author{Nitesh Kumar}\email{nitesh.kumar@bose.res.in}
\affiliation{S. N. Bose National Centre for Basic Sciences, Salt Lake City, Kolkata-700106, India}


\begin{abstract}
In typical rare-earth lanthanide compounds, the localized 4\textit{f}‐electrons have a weak effect on the electrical conduction, limiting their influence on the Berry curvature and, hence, the intrinsic anomalous Hall effect. A comprehensive study of the magnetic, thermodynamic, and transport properties of single-crystalline NdGaSi, guided by first‐principles calculations, reveals a ferromagnetic ground state that induces a splitting of quasi‐flat 4\textit{f} electronic bands and positions them near the Fermi energy. The observation of an extraordinarily large intrinsic anomalous Hall conductivity of 1165 $\Omega^{-1}$ cm$^{-1}$ implies the direct involvement of localized states in the generation of non-trivial band crossings around the Fermi energy. The angle-resolved photoemission spectroscopy measurements provide direct evidence of non-trivial crossing of the 4\textit{f}-bands with dispersive bands. These results are remarkable when compared to ferrimagnetic NdAlSi, which differs only in a non-magnetic atom (a change in the principal quantum number \textit{n} of the outer \textit{p }orbital) with the same number of valence electrons and does not exhibit any measurable anomalous Hall conductivity.

\end{abstract}

\maketitle

\vspace{3mm}

4\textit{f}-electrons are typically confined to the ion core and strongly shielded by the more delocalized 5\textit{d} and 5\textit{p} orbitals \cite{johansson1979energy}. They, therefore, rarely contribute to electrical conduction, unlike their itinerant 3\textit{d}-electron counterparts. Nevertheless, flat bands of 4\textit{f}-electrons carry robust local magnetic moments and exhibit significantly enhanced spin-orbit coupling (SOC), which can theoretically generate large Berry curvatures (BC) via strong hybridization with dispersive bands and, hence, a substantial intrinsic anomalous Hall conductivity (AHC). However, the literature on the anomalous Hall effect (AHE) in rare-earth-based compounds is scarce, while the largest values have been reported in 3\textit{d}-element-based itinerant magnets. The dual role of conduction and magnetism by 3\textit{d}-electrons ensures that SOC strongly modulates the electronic band structure, generating a substantial BC that deflects carriers transversely, leading to a large intrinsic contribution to the AHC \cite{liu2018giant,nayak2016large,nakatsuji2015large,guin20212d}. On the other hand, in 4\textit{f}-compounds, in general, the localized 4\textit{f}-electrons only contribute to magnetism, while conduction is provided by more dispersive bands that don't constitute the \textit{f}-electrons  \cite{ yang2021noncollinear,alam2023sign,shekhar2018anomalous,dhital2023NAG, kotegawa2024large}. This decoupling dramatically reduces the amplitude of the BC in the conduction bands.

However, it has been argued that quasi-flat bands enhance the BC effects by effectively increasing the momentum-space separation between Weyl crossings resulting from time-reversal symmetry breaking \cite{zhou2019weyl,jiang2021giant,Yao2024PRBFB}. In addition, if the quasi-flat bands coexist with dispersive bands just below the Fermi energy (\textit{E}$_\mathrm{F}$), the situation is prone to provide non-trivial crossing points, thereby enhancing the BC. effect in the presence of SOC.  This establishes a basis for the combined
experimental and theoretical study on the effect of quasiflat
bands of 4\textit{f} electrons on the BC and consequently the AHC. In this context, we have studied NdGaSi, which belongs to a large family of compounds with tetragonal symmetry \cite{gaudet2021weyl,chang2018magnetic,suzuki2019singular,yang2021noncollinear,puphal2020topological,lyu2020nonsaturating,yao2023large,yamada2024nernst,wang2022ndalsi,wang2023quantum,piva2023topological,ram2023magnetic,gong2024anomalous,zhang2024magnetism}. It has been shown that substitution in the Ga or Si site can drastically affect the magnetic exchange mechanism, resulting in compounds exhibiting a variety of magnetic ground states such as ferrimagnetism (FIM), antiferromagnetism (AFM), helical states, and ferromagnetism (FM) \cite{bouaziz2024origin}. We show that the magnetic ground state is closely related to the position of the quasi-flat\textit{ f}-bands, i.e., whether they form the valence band or remain unoccupied. In the case of NdGaSi, we show from magneto-transport, thermodynamic measurements, guided by first-principles calculations, that the FM ground state ensures that the quasi-flat \textit{f}-bands lie in the vicinity of \textit{E}$_\mathrm{F}$. This we have directly probed by angle-resolved photoemission spectroscopy (ARPES), and is further corroborated by the large enhancement of the density of states (DOS) at \textit{E}$_\mathrm{F}$ and specific heat measurements, resulting in an unusually large Sommerfeld coefficient. This ultimately leads to the observation of an extraordinarily large AHC due to the BC effect, in contrast to FIM NdAlSi, where the\textit{ f}-bands are pushed to the unoccupied states and therefore show vanishingly small AHC \cite{wang2022ndalsi,gaudet2021weyl}.

\begin{figure}[t]
\centering
\includegraphics[width=0.45\textwidth]{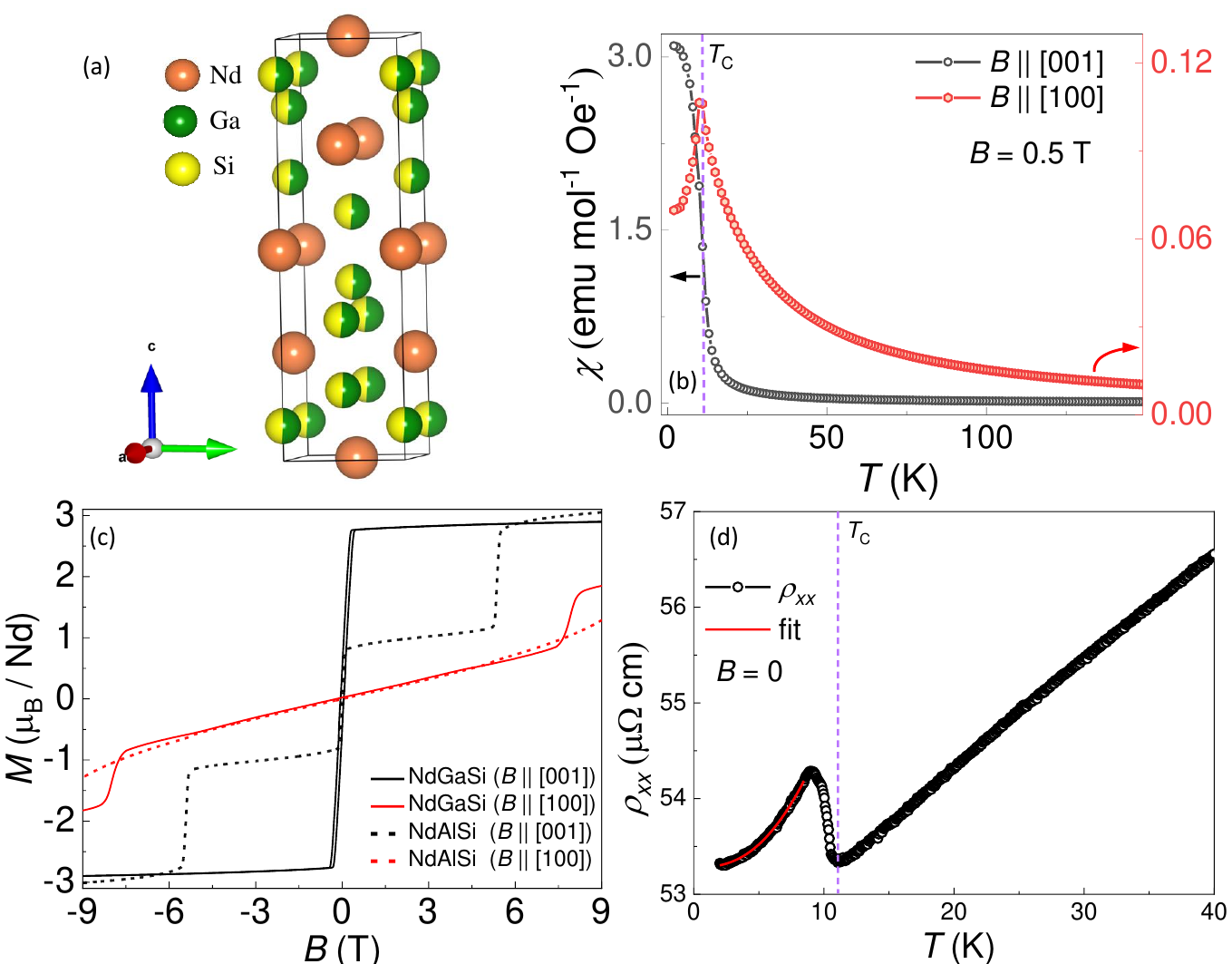}
\caption{(a) A tetragonal unit cell of NdGaSi. A simultaneous occupancy of Ga and Si sites (complete site mixing) imparts a centrosymmetric space group \textit{I}4$_1$\textit{/amd}. (b) Magnetic susceptibility ($\chi$)  measured under the field-cooled (FC) condition by applying the magnetic field along the in-plane and out-of-the-plane applied magnetic field. (c) Isothermal magnetization curves of NdAlSi and NdGaSi at 2 K with the magnetic field applied along [001] and [100]. (d) Longitudinal resistivity ($\rho_{xx}$) as a function of temperature under zero magnetic field, along with a fit (red) at low temperature to obtain the coefficient of electron scattering. }
\label{fig1}
\end{figure}

We have successfully synthesized high-quality single crystals of NdGaSi, which crystallize in the tetragonal $\alpha$-ThSi$_2$-type centrosymmetric structure (space group {\textit{I}}4$_1$\textit{/amd}). Both the single-crystal x-ray diffraction and neutron diffraction studies confirm that the non-rare-earth sites are equally occupied by Ga and Si, forming well-ordered repeating stacks of Nd and Ga/Si. The consecutive Nd-layers are separated by a distance of 3.57 Å, creating a distinctive layered arrangement (Fig. \ref{fig1}a). The detailed methodology for crystal growth, characterization, and structure determination is provided in Sec. IA of the supplementary information (SI) \cite{supply}.

The field-cooled magnetic susceptibility (Fig. \ref{fig1}b) along the [001] direction is nearly 44 times larger than along the [100] direction, indicating strong magnetic anisotropy with the out-of-plane direction [001] as the easy axis. This striking difference is a common feature of other compounds in this series \cite{yang2021noncollinear,lyu2020nonsaturating,wang2022ndalsi}. A clear anomaly at around \textit{T}$\mathrm{_C}$ = 11 K for both directions confirms the onset of magnetic ordering.  The out-of-plane susceptibility shows a sudden rise in magnetization with decreasing temperature, reminiscent of an FM. In contrast,  the in-plane susceptibility reveals a cusp-like feature, with a peak at \textit{T}$\mathrm{_C}$, which is a feature of strong anisotropic uniaxial ferromagnets \cite{suter1984transverse,hudl2014thermodynamics,xu2024giant}.

FM in NdGaSi is further indicated by a sharp and spontaneous saturation of magnetization to a value of 2.9 $\mu_\mathrm{B}$/Nd when the magnetic field is applied along [001] in an isothermal magnetization measurement at 2 K. The corresponding magnetization along [100] increases monotonically with increasing magnetic field up to 9 T, except at 7.9 T, where a metamagnetic transition occurs. The magnetic behavior in NdGaSi is quite different from its sister compound NdAlSi, which chemically differs only in the nonmagnetic Ga/Al site occupancy. NdAlSi is known to exhibit FIM nature with a spontaneous magnetization of 1 $\mu_\mathrm{B}$/Nd when the magnetic field is applied along \textit{c}-axis. The full polarization of the Nd-moment, i.e., 3 $\mu_\mathrm{B}$/Nd, is only achieved after a magnetic field-induced meta-magnetic transition at 5 T (Fig. \ref{fig1}c).

NdGaSi exhibits metallic behavior down to 2 K (Fig. \ref{fig1}d), with a notable anomaly around 11 K in the longitudinal resistivity $\rho_{xx}$, consistent with the magnetic transition. The low-temperature curve below the magnetic ordering is best expressed by the electron-electron scattering term (2nd term) along with a gapped magnon contribution (3rd term) as $\rho_{xx} = \rho_0 + AT^2 +  CT^2e^{-\Delta/T}$, where $\rho_0$ represents the scattering of electrons from lattice imperfections \cite{mott1936h,kaczorowski2014magnetic,ram2024crystalline}. We obtain \textit{A} = 0.624 x 10$^{-2}$ $\mu \Omega $ cm K$^{-2}$ and $\Delta =$ 7.57 K from the fit. NdGaSi exhibits a relatively small residual resistivity ratio (RRR) of $\sim$ 1.66, which could be the result of anti-site disorder between Si and Ga atoms. We will discuss the magnitude of \textit{A} in the next section under the context of electron localization.

\begin{figure}[t]
\centering
\includegraphics[width=0.48\textwidth]{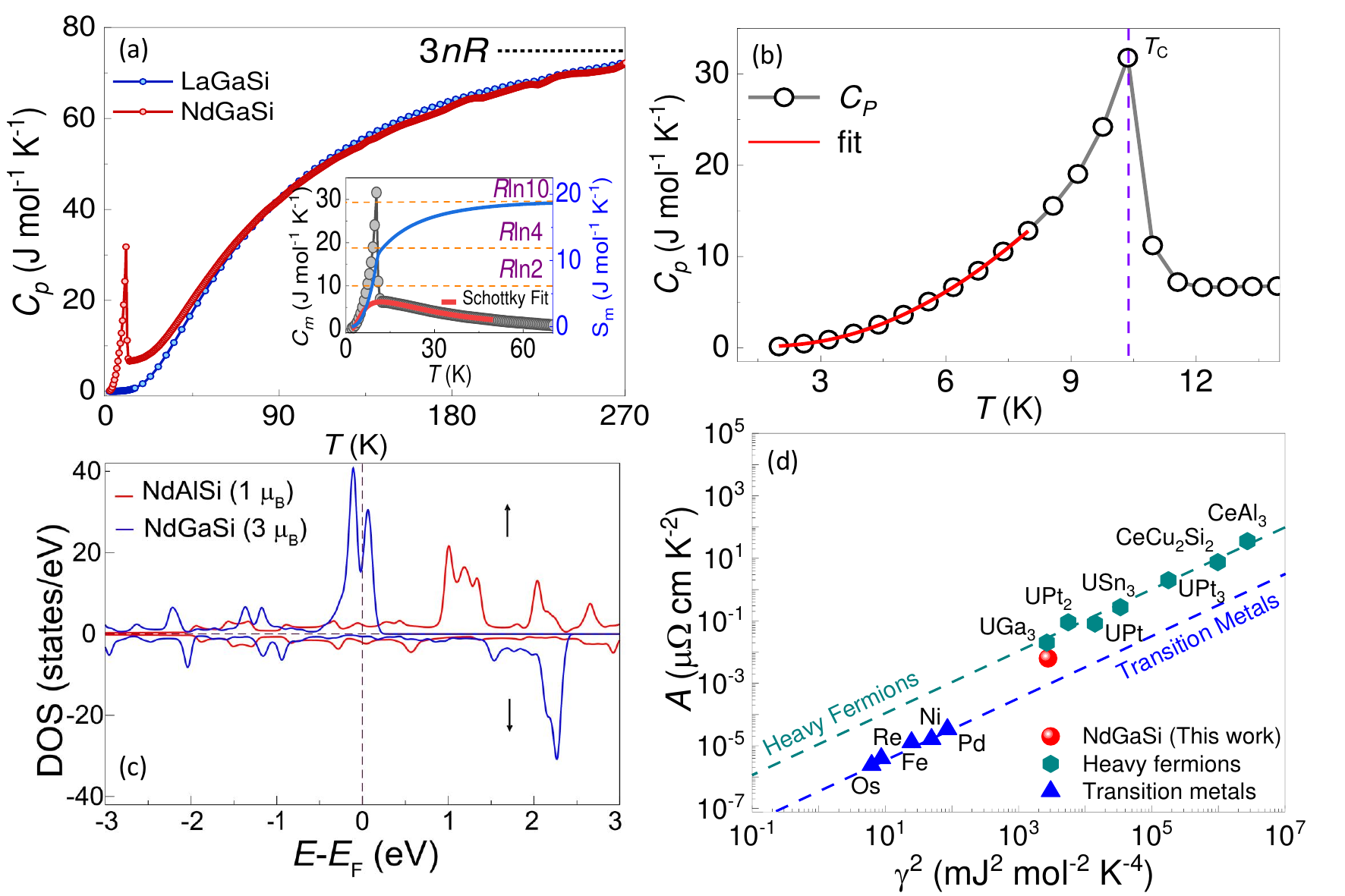}
\caption{(a) Specific heat of NdGaSi and NdAlSi as a function of temperature. The inset shows the variation of both magnetic specific heat and entropy as a function of temperature. (b) Specific heat data below 12 K with fitting. (c) Density of states of NdGaSi and NdAlSi in the ground state. (d) Kadowaki-Woods ratio of NdGaSi and its representation among itinerant and heavy Fermion systems. }
\label{fig2}
\end{figure}

Analysis of low-temperature specific heat provides essential clues regarding localized states around \textit{E$_\mathrm{F}$}. NdGaSi exhibits a sharp $\lambda$-like peak around 11 K corresponding to the bulk magnetic transition (Fig. \ref{fig2}a). A broad hump centered around 15 K indicates the well-known Schottky anomaly, which arises due to the splitting of the Nd$^{3+}$ atomic levels induced by crystalline electric field (CEF) \cite{waltercef,gopal2012specific}.  The magnetic specific heat ($S_m(T) = \int\frac{C_m}{T}dT$)  attains a value of  $\sim$\textit{ R}ln2 below\textit{ T}$\mathrm{_C}$ (See inset of Fig.  \ref{fig2}a), indicating the localized nature of 4\textit{f}-electrons. Furthermore,  the specific heat data below\textit{ T}$\mathrm{_C}$ can be expressed as $C_p(T) = \gamma T + \beta T^3 + \delta T^{3/2}e^{-\Delta/T}$, where the first, second, and third terms correspond to the electron, lattice, and magnon contributions, respectively \cite{gopal2012specific,kaczorowski2014magnetic}. The dominant exponential behavior of $C_P (T)$ at a low temperature indicates the presence of a spin gap that arises due to anisotropic FM exchange, and fitting was carried out by fixing the lattice-related parameter obtained from that for LaGaSi in Sec. IV of SI \cite{supply}, which gives a value of the magnon gap ($\Delta \sim$ 6.8 K), consistent with that obtained from resistivity analysis.  However, we find that the obtained Sommerfield coefficient ($\gamma\sim 55.46 $ mJ mol$^{-1}$ K$^{-2}$)  is relatively high compared to previously known Nd compounds \cite{campoy2006magnetoresistivity,szytula2007electronic,ram2024crystalline} and strays away from typical itinerant systems. On the other hand, LaGaSi, which does not contain\textit{ f}-electrons, exhibits a reasonably small $\gamma$ of 2.54 mJ mol\textsuperscript{-1 }K\textsuperscript{-2}. This suggests that the enhancement of DOS is mainly contributed by the \textit{f}-electrons in NdGaSi.

The Sommerfeld coefficient ($\gamma$) dictates the distribution of the DOS around $E_\mathrm{F}$. For a free electron model, the theoretical value of $\gamma$  $(\gamma_{theo})$ is $ D(E_\mathrm{F})\pi^2k_B^2/3$, where $D(E_\mathrm{F})$ is the DOS at \textit{E$_\mathrm{F}$}, and if we assume that all the DFT DOS comes from the free electron, then the theoretical value ($\gamma_{theo}$) is $\sim $ 43.39 mJ mol$^{-1}$ K$^{-2}$, significantly smaller than the experimentally observed value $\gamma$. The enhanced value of $\gamma$ suggests the presence of correlated electrons at \textit{$E_\mathrm{F}$} \cite{grimvall1976electron,xiao2024preparation}, which is consistent with the presence of \textit{f}-states at\textit{ $E_\mathrm{F}$} shown in Fig. \ref{fig2}c.

Our calculated spin-resolved DOS of NdGaSi unveils localized \textit{f}-states at \textit{E}$_\mathrm{F}$. It reveals an FM ground state, where Nd 4\textit{f}-electrons are the primary source of magnetism with a magnetic moment of 3 $\mu_\mathrm{B}$ per formula unit. As illustrated in Fig.  \ref{fig2}c (blue line), the spin-resolved DOS exhibits a tiny gap in the minority spin channel, while the majority spin channel remains metallic, indicating a nearly half-metallic behavior. In contrast, NdAlSi exhibits an FIM ground state with a low magnetic moment of 1 $\mu_\mathrm{B}$ per formula unit, which can be understood in terms of the DOS, represented by a red curve in Fig.  \ref{fig2}c.  A small value of DOS at \textit{E}$_\mathrm{F}$ corresponds to dispersive bands, which mostly comprise the Al and Si states (see Sec. VII of SI \cite{supply}). The large magnetic moment (3 $\mu_\mathrm{B}$) of NdGaSi can be achieved by depopulating the localized up-spin \textit{f}-bands close to \textit{E}$_\mathrm{F}$ that give rise to a significant peak in the DOS around the \textit{E}$_\mathrm{F}$ (see blue curve in Fig.  \ref{fig2}c). A significant peak in DOS arising from the localized up-spin \textit{f}-electrons is also observed in EuAlSi \cite{bouaziz2024origin}, which exhibits an FM ground state.

The weaker hybridization strength of the 4\textit{f}–4\textit{p} (NdGaSi) compared to the 4\textit{f}–3\textit{p} (NdAlSi) along the \textit{c}-axis plays a crucial role in magnetism  \cite{maletta1988itinerant,andreev1998magnetic}. The super-exchange mechanism is highly favored in NdAlSi due to the strong 4\textit{f}–3\textit{p} hybridization, which discourages the formation of a local Nd moment and strongly favors a short-range super-exchange behavior, evident from FIM behavior. NdGaSi, on the other hand, hosts a weaker 4\textit{f}–4\textit{p} hybridization than the former and strongly promotes a long-range 4\textit{f}–4\textit{f}  Ruderman-Kittel-Kasuya-Yosida (RKKY) interaction. This is highly favored by the majority of localized \textit{f}-states around the \textit{E}$_\mathrm{F}$ realized earlier, providing enough conducting states to support an RKKY interaction \cite{bouaziz2024origin}. The RKKY interaction seems to promote FM behaviour in NdGaSi, which is evident from our experimental results.
\begin{figure*}
\centering
\includegraphics[width=0.8\textwidth]{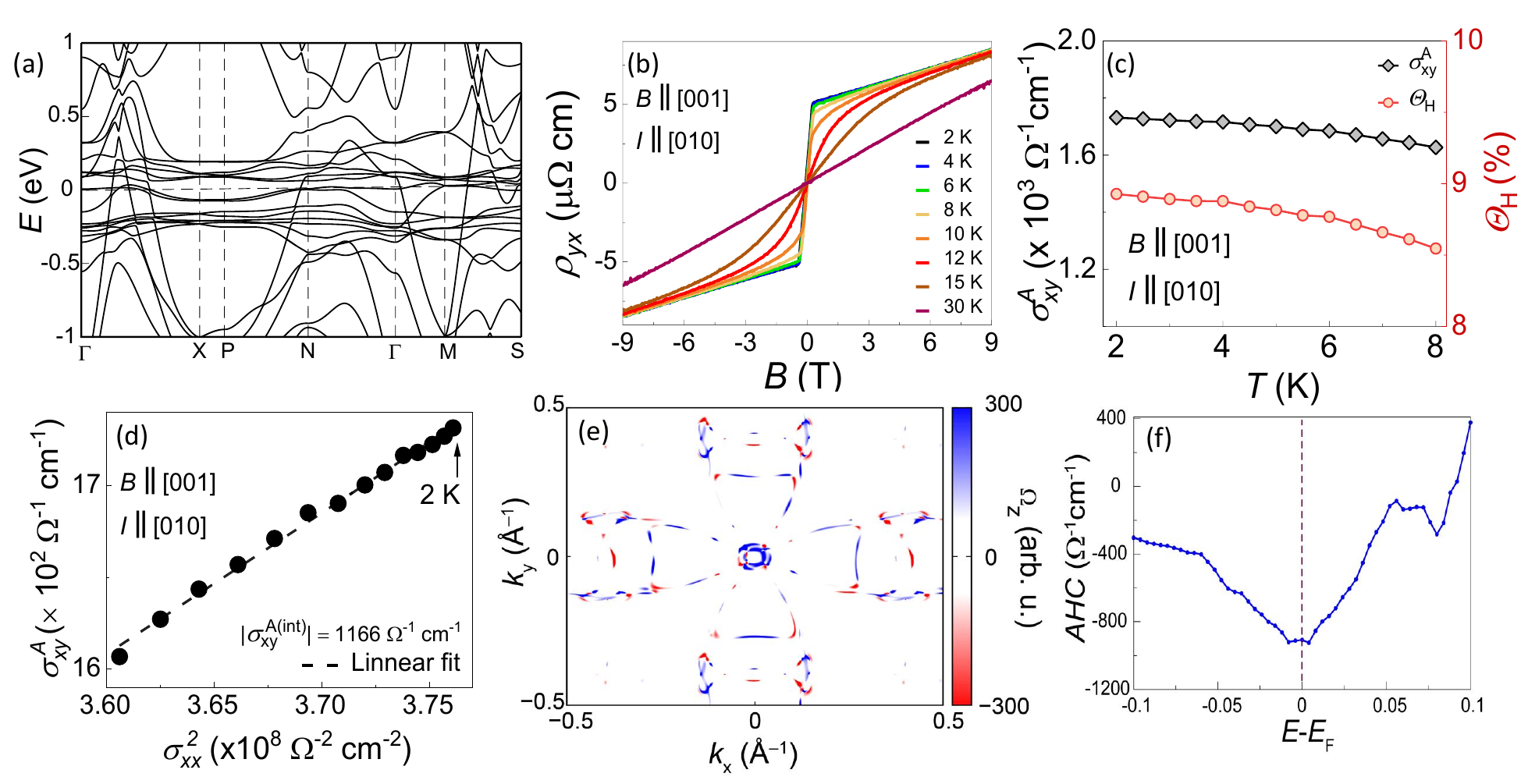}
\caption{(a) The band structure of NdGaSi with SOC in the FM state, with the spins aligned along [001]. (b) Magnetic field-dependent Hall resistivity ($\rho_{yx}$) at different temperatures ranging from 2 K to 30 K with $B \parallel$ [001] and $I \parallel$ [010]. (c) Temperature dependence of the anomalous Hall conductivity ($\sigma^{A}_{xy}$) and anomalous Hall angle ($\Theta_\mathrm{H}$).  (d) Linear fitting of anomalous Hall conductivity $\sigma^{A}_{xy}$ vs. $\sigma^{2}_{xx}$ curve. (e) BC distribution of NdGaSi in the first Brillouin zone reveals regions of maximum positive and negative values, depicted in red and blue, respectively. (f) Energy-dependent AHC obtained from the first-principles calculations.} 
\label{fig3}
\vspace{-0.5cm}
\end{figure*}

To conclude our argument regarding the occupied localized states, we shift our focus to the Kadowaki-Woods ratio relating $\gamma$ to the coefficient of the \textit{T}$^2$ term in the longitudinal resistivity, \textit{A}, i.e., $R_\mathrm{KW} = A/\gamma^2$. This is used as a benchmark for electron localization, with a universal constant for heavy fermion systems ($R_\mathrm{KW} \sim 10^{-5} \mu \Omega$ cm K$^2$ mol$^2$ mJ$^{-2}$) and transition elements ($R_\mathrm{KW} \sim 10^{-7} \mu \Omega$ cm K$^2$ mol$^2$ mJ$^{-2}$) \cite{kadowaki1986universal,rice1968electron,jacko2009unified}. NdGaSi yields a value of $R_\mathrm{KW} \sim$ 0.225 x $10^{-5} \mu \Omega$ cm K$^2$ mol$^2$ mJ$^{-2}$, approximately two orders larger than itinerant systems but slightly smaller than conventional localized systems. This is depicted in Fig. \ref{fig2}d, where it lies in a regime best defined by moderately enhanced carrier localization but not to the extent of heavy-fermion systems.

The calculated electronic band structure of NdGaSi with and without SOC confirms localized \textit{f}-electrons in a flat band near \textit{E$_\mathrm{F}$}, and the details of electronic bands without SOC are provided in Sec. VII of SI \cite{supply}. Our calculations, in the presence of SOC and the magnetic spin aligned along [001], show that crossing points appear along $\Gamma$–X and $\Gamma$–M due to the interaction between Nd \textit{f}-orbitals and Ga \textit{p}-orbitals, as shown in Fig. \ref{fig3}a. While the\textit{ f}-bands remain largely flat, they exhibit weak dispersion near these crossings, forming non-trivial nodal points. We have also calculated the Berry curvature along the same k-path, which further confirms the nontrivial nature of the identified nodal points, as shown in Sec. VII of the SI \cite{supply}.

The inclusion of localized \textit{f}-electrons as occupied states can significantly impact the BC and, consequently, lead to a large AHC in NdGaSi. It is noteworthy that NdAlSi, in which the localized states lie well above the \textit{E}$_\mathrm{F}$, does not exhibit any measurable AHC \cite{wang2022ndalsi,gaudet2021weyl}. Fig. \ref{fig3}b represents the field-dependent Hall resistivity ($\rho_{yx}$) curve measured with $I \parallel $ [010] and $B \parallel $ [001] at various temperatures. At 30 K ($> T_\mathrm{C}$), $\rho_{yx}$ is linear with increasing magnetic field, as expected for a metallic system with carrier concentration (\textit{n}) $\sim$  9.2 x 10$^{22}$ cm$^{-3}$. Below $T_\mathrm{C}$, the anomalous Hall resistivity ($\rho_{yx}^A$) becomes prominent above a minuscule magnetic field of $\sim$ 0.3 T and exhibits a linear trend up to 9 T, mimicking the magnetization isotherms \cite{supply}. Conventionally, the total Hall resistivity in a FM can be represented as $\rho_{yx} (B)$ = $\rho_{yx} ^O$ + $\rho_{yx} ^A$ = $R_0B$ + $R_s$$\mu_0$$M$, where $\rho_{yx} ^O$ and $\rho_{yx} ^A$ are the ordinary and anomalous contributions to the total Hall resistivity, with $R_0$ and $R_s$ being the ordinary and anomalous Hall coefficients, respectively \cite{nagaosa2010anomalous,karplus1954hall,jungwirth2002anomalous}.  Interestingly, we also observe a finite anomalous Hall effect even above $T_\mathrm{C}$. This could originate from either short-range magnetic ordering or from the magnetic field-induced time reversal symmetry breaking in the presence of strong magnetic fluctuations near $T_\mathrm{C}$ \cite{ma2019eca,forslund2025anomalous}.  We discuss its possible origin in the context of NdGaSi in the SI \cite{supply} (Section VIIID).

The Hall conductivity ($\sigma_{xy}$) is determined from the tensor relation $\sigma_{xy} = \rho_{yx}/(\rho_{xx}^2 + \rho_{yx}^2)$ as a function of the magnetic field \cite{supply}, where $\rho_{yx} = -\rho_{xy}$ and the expressions $\rho_{xx}^2 = \rho_{xx}\rho_{yy}$ and $\rho_{xx} = \rho_{yy}$ originate from the in-plane isotropy of the tetragonal crystal structure of NdGaSi. Using zero-field extrapolation of high-field $\sigma_{xy}$ data on the \textit{y}-axis, we have extracted the AHC, denoted by $\sigma_{xy}^A$. Remarkably, $\sigma_{xy}^A$ $\sim$ 1730 $\Omega^{-1}$ cm$^{-1}$ at 2 K, with a subtle change up to 8 K in the ordered regime (Fig. \ref{fig3}c), among the largest in 4\textit{f}-electron systems and comparable to the largest known itinerant 3\textit{d} systems to the best of our knowledge \cite{shekhar2018anomalous,arai2024intrinsic,liu2018giant,meng2019large,nayak2016large,nakatsuji2015large,wang2017anisotropic}. We have noticed a large anomalous Hall angle (AHA) of $\sim$ 9.3 \% at 2 K (Fig. \ref{fig3}c). 

AHE in NdGaSi occurs at very low temperatures ($<$ 11 K), nullifying the role of phonon scattering and suggesting domination by residual resistivity ($\rho_{xx0}$). In such a scenario, $\sigma_{xy}^A$ can be scaled using the empirical formula $\sigma_{xy}^A = (\alpha\sigma_{xx0}^{-1} + \beta\sigma_{xx0}^{-2})\sigma_{xx}^2 + m$, where $\sigma_{xx0}$ is the residual longitudinal conductivity and the coefficients $\alpha$, $\beta$, and $m$ correspond to the skew-scattering, side jump, and the intrinsic Berry phase contributions to $\sigma_{xy}^A$, respectively \cite{tian2009proper,hou2015multivariable}. The \textit{m} (or $\sigma_{xy}^{A, int}$) is determined by employing a linear fitting between $\sigma_{xy}^A$ and $\sigma_{xx}^2$ as plotted in Fig. \ref{fig3}d. The  $\sigma_{xy}^{A, int}$, thus, obtained is equal to 1166 $\Omega^{-1}$ cm$^{-1}$, further suggesting that the intrinsic Berry phase mechanism dominates the AHE in NdGaSi, however, the contribution of the scattering-dependent terms cannot be ignored. If one considers that each Nd layer independently contributes a quantum conductance of $e^2/h$, then the total BC-related AHC can be estimated to be  $e^2/hd\sim$ 1080 $\Omega^{-1}$ cm$^{-1}$, where \textit{e} is the electronic charge, \textit{h} is the Planck constant, and \textit{d} is the inter-layer spacing between two consecutive Nd layers \cite{xiao2010berry}. This is quite close to our estimation of $\sigma_{xy}^{A, int} $ = 1166 $\Omega^{-1}$ cm$^{-1}$ from the scaling analysis. Finally, to confirm the reproducibility, we have measured $\sigma_{xy}^A$ at 2 K with $B \parallel $ [001] and $I \parallel $ [010] for several NdGaSi single crystals, as described in Sec. VIIIE of SI \cite{supply}. We have also compared NdGaSi with various other 4\textit{f}-electron-based compounds in terms of intrinsic AHC and their possible origin in Table S8 of the SI \cite{supply}. 
It can be seen that AHC in these systems almost always originates from the crossings of \textit{s}, \textit{p}, and \textit{d}-bands, and therefore, quasi-flat 4\textit{f}-bands-mediated AHC, as observed in NdGaSi, is rare.

The large intrinsic AHC in experiments must arise from the BC associated with non-trivial crossings in the electronic band structure below \textit{E}$_\mathrm{F}$ in the presence of SOC. $\sigma_{xy}^{A,int}$ can be calculated using the Kubo formalism within the framework of linear response theory~\cite{gradhand2012first} and it is written as a function of $z$-component of BC ($\Omega_z$), with further details provided in Sec VII of SI \cite{supply}. $\Omega_z$ is shown in the \textit{k}$_x$–\textit{k}$_y$ plane for $k_z=0$ in first Brillouin zone (see Fig. \ref{fig3}e).  Fig. \ref{fig3}f presents the energy-dependent AHC, highlighting a significant contribution near\textit{ E}$_\mathrm{F}$. Our calculations predict an AHC magnitude of 934 $\Omega^{-1}$ cm$^{-1}$ , which is in excellent agreement with our experimental measurements for the material. Small carrier doping, i.e, small change in $E_\mathrm{F}$, shows a weak effect on the AHC as evident from Fig. \ref{fig3}f.
 
\begin{figure}[t]
\centering
\includegraphics[width=0.49\textwidth]{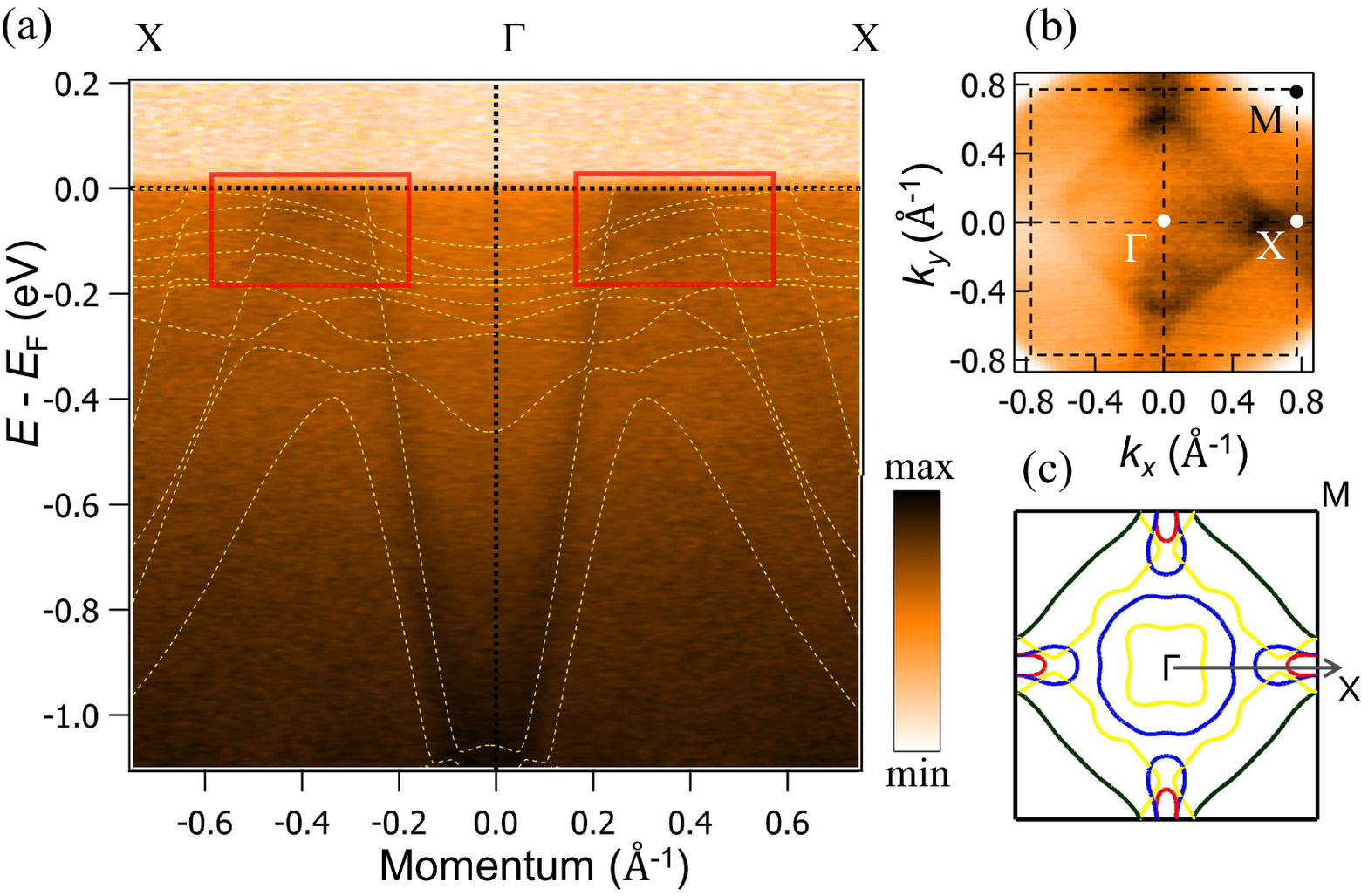}
\caption{(a) ARPES spectra of NdGaSi along $\Gamma -\mathrm{X}$ direction with incident photon energy of 120 eV recorded using linear horizontal polarized light. The spectra are presented with the overlapped band structure obtained from DFT (with SOC) represented by dotted yellow lines. Red boxes indicate the quasiflat bands observed from ARPES.  (b) Experimental Fermi surface recorded using \textit{hv} = 50 eV and linear polarized light. (c)  Theoretical Fermi surface calculated from DFT at the $\Gamma-\mathrm{X}-\mathrm{M}$ plane ($k_z  = 0$).}
\label{fig4}
\end{figure}
The Nd-derived 4\textit{f-}bands exhibit anticrossing with the dispersive bands, and their hybridization is the primary source of the large BC observed in our calculations. Our ARPES measurements corroborate this claim, overlapped with the DFT band structure along $\mathrm{X}-\Gamma-\mathrm{X}$, as shown in Fig. \ref{fig4}a. The ARPES spectrum reveals a signature of the flat bands crossing with the dispersive band with enhanced spectral weight. Fig. \ref{fig4}b displays the experimentally measured Fermi surface (FS), while Fig. \ref{fig4}c shows the one in the $\Gamma-\mathrm{X}-\mathrm{M}$ plane, reproducing the fourfold symmetry of the Brillouin zone. A compact, nearly square-shaped pocket at $\Gamma$ (yellow) originates from a dispersive band, while a surrounding $\Gamma$-centered sheet (blue) arises from a part of the quasi-flat band, as identified in the SI band structure \cite{supply}. At $\mathrm{X}$ points, elongated and lobe-shaped pockets (blue), along with an additional smaller lobe-shaped pocket (red), also originate from the quasi-flat \textit{f}-bands highlighted in the SI (See Fig. S7a) \cite{supply}. The calculated FS shows good agreement with ARPES FS, including the pronounced intensity shown near the $\mathrm{X}$ point in Fig. \ref{fig4}b.

In summary, we observe a complex interplay between magnetism that arises from localized \textit{f}-electrons and topology in the electronic band structure in NdGaSi.  The enhanced Sommerfeld coefficient and the Kadowaki-Woods ratio underscore moderate electron correlations and enhanced carrier localization of conduction electrons due to the presence of flat bands at\textit{ E}$_\mathrm{F}$. Consequently, a substantial BC arises from emerging Weyl points generated by the crossing of the \textit{f}-bands with dispersive bands. Our ARPES data are in good agreement with DFT calculations and suggest an interaction between \textit{f} states and the dispersive bands, resulting in additional spectral weight near the $E_\mathrm{F}$. This corroborates the experimentally observed giant intrinsic AHC. We propose that the manipulation of the magnetic ground state provides a straightforward yet powerful tool to fully exploit the flat bands. By finely tuning the magnetic interactions to achieve just a sufficient magnitude of band splitting, one can relocate the localized states at \textit{E}$_\mathrm{F}$ and harness their exotic properties. Our work provides a robust framework for utilizing localized \textit{f}-bands to control large AHE in rare-earth intermetallic compounds and pave a pathway for engineering the band structure in real materials, enabling the exploration of these otherwise elusive states.

\vspace{3mm}

\section*{ACKNOWLEDGEMENTS}

NK acknowledges DST for financial support through Grant Sanction No. CRG/2021/002747 and Max Planck Society for funding under the Max Planck-India partner group project. This research project made use of the instrumentation facility provided by the Technical Research Centre (TRC) at the S.N. Bose National Centre for Basic Sciences, under the Department of Science and Technology, Government of India. JS and MK acknowledge Ayan Jana for his assistance. JS and MK also acknowledge the National Supercomputing Mission (NSM) for providing computing resources of ‘PARAM RUDRA’ at S.N. Bose National Centre for Basic Sciences, which is implemented by C-DAC and supported by the Ministry of Electronics and Information Technology (MeitY) and Department of Science and Technology (DST), Government of India. VM, RL, and AF acknowledge Deutsche Forschungsgemeinschaft for Grant SFB 1143 (project C04), the Wuerzburg-Dresden Cluster of Excellence on Complexity and Topology in Quantum Matter – ct.qmat (EXC 2147, projectID 390858490). S.T. thanks the Science and Engineering Research Board (SERB), India, for the financial support through Grant No. CRG/2023/00748.

\makeatletter
\providecommand \@ifxundefined [1]{%
 \@ifx{#1\undefined}
}%
\providecommand \@ifnum [1]{%
 \ifnum #1\expandafter \@firstoftwo
 \else \expandafter \@secondoftwo
 \fi
}%
\providecommand \@ifx [1]{%
 \ifx #1\expandafter \@firstoftwo
 \else \expandafter \@secondoftwo
 \fi
}%
\providecommand \natexlab [1]{#1}%
\providecommand \enquote  [1]{``#1''}%
\providecommand \bibnamefont  [1]{#1}%
\providecommand \bibfnamefont [1]{#1}%
\providecommand \citenamefont [1]{#1}%
\providecommand \@href[1]{\@@startlink{#1}\@@href}%
\providecommand \@@href[1]{\endgroup#1\@@endlink}%
\providecommand \@sanitize@url [0]{\catcode `\\12\catcode `\$12\catcode `\&12\catcode `\#12\catcode `\^12\catcode `\_12\catcode `\%12\relax}%
\providecommand \@@startlink[1]{}%
\providecommand \@@endlink[0]{}%
\providecommand \@url [1]{\endgroup\@href {#1}{\urlprefix }}%
\providecommand \urlprefix  [0]{URL }%
\providecommand \doibase [0]{https://doi.org/}%
\providecommand \selectlanguage [0]{\@gobble}%
\providecommand \bibinfo  [0]{\@secondoftwo}%
\providecommand \bibfield  [0]{\@secondoftwo}%
\providecommand \translation [1]{[#1]}%
\providecommand \BibitemOpen [0]{}%
\providecommand \bibitemStop [0]{}%
\providecommand \bibitemNoStop [0]{.\EOS\space}%
\providecommand \EOS [0]{\spacefactor3000\relax}%
\providecommand \BibitemShut  [1]{\csname bibitem#1\endcsname}%
\let\auto@bib@innerbib\@empty
\bibliography{reference}

\end{document}